\documentclass[aps,prd,amsmath,floats,floatfix,twocolumn,
  superscriptaddress,nofootinbib,showpacs]{revtex4-2}

\usepackage[usenames, dvipsnames]{xcolor}
\usepackage{hyperref}
\hypersetup{
  colorlinks,
  citecolor={cyan!50!black},
  linkcolor={orange!60!black},
  urlcolor=violet,
}
\usepackage{orcidlink}
\usepackage{graphicx}
\usepackage{amssymb}
\usepackage{mathtools}  %
\usepackage{tensor}
\usepackage{appendix}
\usepackage{bm} %
\usepackage{commath}  %
\usepackage{physics}
\usepackage{bbm} %
\usepackage{microtype}
\usepackage{enumitem}
\usepackage{siunitx}
\usepackage{csquotes}
\usepackage[capitalise]{cleveref}
\usepackage[caption=false]{subfig}
\usepackage{tikz}
\usepackage[normalem]{ulem} %

\newcommand{\spec}{\texttt{SpEC}}
\newcommand{\spectre}{\texttt{SpECTRE}}

\begin{document}

\title{Parameter control for binary black hole initial data}

\newcommand{\aei}{\affiliation{Max Planck Institute for Gravitational Physics
(Albert Einstein Institute), Am M{\"u}hlenberg 1, Potsdam 14476, Germany}}
\newcommand{\caltech}{\affiliation{Theoretical Astrophysics, Walter Burke
Institute for Theoretical Physics, California Institute of Technology, Pasadena,
California 91125, USA}}
\newcommand{\oberlin}{\affiliation{Department of Physics and Astronomy, Oberlin
College, Oberlin, Ohio 44074, USA}}

\author{Iago B.~Mendes\,\orcidlink{0009-0007-9845-8448}} \email{imendes@caltech.edu} \caltech
\author{Nils L. Vu\,\orcidlink{0000-0002-5767-3949}} \email{nilsvu@caltech.edu} \caltech
\author{Oliver Long\,\orcidlink{0000-0002-3897-9272}} \aei
\author{Harald P. Pfeiffer\,\orcidlink{0000-0001-9288-519X}} \aei
\author{Robert Owen\,\orcidlink{0000-0002-1511-4532}} \oberlin

\date{\today}

\begin{abstract}
  When numerically solving Einstein's equations for binary black holes (BBH), we
  must find initial data on a three-dimensional spatial slice by solving
  constraint equations.
  The construction of initial data is a multi-step process, in which one
  first chooses freely specifiable data that define a conformal
  background and impose boundary conditions.
  Then, one numerically solves elliptic equations and calculates physical
  properties such as horizon masses, spins, and asymptotic quantities from the
  solution.
  To achieve desired properties, one adjusts the free data in an
  iterative ``control'' loop.
  Previous methods for these iterative adjustments rely on Newtonian
  approximations and do not allow the direct control of total energy and angular
  momentum of the system, which becomes particularly important in the study of
  hyperbolic encounters of black holes.
  Using the \spectre{} code, we present a novel parameter control procedure
  that benefits from Broyden's method in all controlled quantities. We use this
  control scheme to minimize drifts in bound orbits and to enable the
  construction of hyperbolic encounters.
  We see that the activation of off-diagonal terms in the control Jacobian gives
  us better efficiency when compared to the simpler implementation in the
  Spectral Einstein Code (\spec{}).
  We demonstrate robustness of the method across extreme configurations,
  including spin magnitudes up to $\chi = 0.9999$, mass ratios up to $q = 50$,
  and initial separations up to $D_0 = 1000M$.
  Given the open-source nature of \spectre{}, this is the first time a parameter
  control scheme for constructing bound and unbound BBH initial data is
  available to the numerical-relativity community.
\end{abstract}

\maketitle

\section{Introduction}\label{sec:intro}

Gravitational-wave (GW) observations of binary black holes (BBH) have played an
essential role in the exploration of gravity in recent years
\cite{LIGOScientific:2016aoc,LIGOScientific:2018mvr,LIGOScientific:2020ibl,2111.03606,1602.03841,1903.04467,2010.14529,2112.06861}.
The search for GW signals, their analysis, and tests of general relativity
all rely on precise knowledge of the expected GW signals emitted by BBH systems.
Numerical relativity (NR) has emerged as a key method to provide waveform
information for GW astronomy.
To get these predicted waveforms, NR codes solve
Einstein's equations using the 3+1 formalism (see \cite{BaumgarteShapiro} for a
review), which splits spacetime into spacelike slices of constant time
coordinate. This involves two major steps: finding initial data that
satisfy the Einstein constraint equations on a first slice, and then
evolving the constraint-satisfying fields to get the future slices. Here, we
focus on the former, known as the initial data (ID) problem.

The initial data of an NR simulation determines the initial physical properties
of the BBH system, such as its masses, spins, and orbital
eccentricity. Precise control over these physical parameters is essential
to perform targeted simulation campaigns over the parameter space, e.g.\ to
construct surrogate models~\cite{1905.09300,2204.01972,1705.07089} and to follow up on
observed gravitational wave events~\cite{1607.05377,1712.05836,1911.02693,2203.10109}.
This is especially relevant as the NR
community prepares for the next generation of gravitational-wave detectors, such
as LISA \cite{LISAConsortiumWaveformWorkingGroup:2023arg}, Cosmic Explorer
\cite{Reitze:2019iox}, and Einstein Telescope \cite{Abac:2025saz}. Among other challenges,
these detectors will measure gravitational-wave events with unprecedented
sensitivity, requiring us to accurately model and subtract loud signals from the
data to reveal fainter signals underneath.

Another case that has gained significant interest recently is NR simulations of
BBH hyperbolic encounters
\cite{Damour:2014afa,Damour:2022ybd,Hopper:2022rwo,Rettegno:2023ghr,Albanesi:2024xus,Swain:2024ngs,Long:2025nmj,Fontbute:2025vdv,Pretorius:2007jn,Sperhake:2008ga,Witek:2010xi,Sperhake:2010uv,Sperhake:2012me,Sperhake:2015siy,Kankani:2024may,Nelson:2019czq,Jaraba:2021ces,Rodriguez-Monteverde:2024tnt,Bae:2023sww,Fontbute:2024amb}.
The asymptotic nature of these systems allows for unambiguous parameterizations
of the initial data, which can be used to compare well-defined observables with
other NR codes as well as perturbative calculations
\cite{Rettegno:2023ghr,Buonanno:2024byg,Swain:2024ngs,Long:2025nmj}. However,
these observables are highly sensitive to the initial data parameters, which
makes it important to maintain accurate control over them.

To find initial data, we use the extended conformal thin-sandwich (XCTS)
formulation of the Einstein constraint equations~\cite{York:1998hy,Pfeiffer:2002iy}.
Before solving the XCTS system
of five elliptic partial differential equations (PDEs), we must choose freely
specifiable data that are used for constructing a conformal background and for
imposing boundary conditions. After solving the XCTS equations, we can measure
physical parameters, such as the horizon masses, spins, and the total
momenta of the system. Typically, we want to choose these physical quantities
before running a BBH simulation, but they can only be measured after numerically
solving the XCTS equations.
Therefore, an iterative control scheme is necessary to adjust the free data in
the XCTS equations such that the desired physical parameters are achieved.

It is important to note that ID control comes in two stages: (1) achieving
physical parameters that can be determined at $t=0$, immediately after solving
the XCTS equations, and (2) achieving physical parameters such as eccentricity
that require brief exploratory evolutions.
This paper is concerned with the first stage, and other works on
eccentricity control~\cite{Knapp:2024yww,Nee:2025zdy} and eccentricity
reduction~\cite{Habib:2025} schemes are concerned with the second stage.

Such a control scheme was implemented in Ref.~\cite{Buchman2012-ud}
and improved in Ref.~\cite{Ossokine:2015yla} in the
Spectral Einstein Code (\spec{})~\cite{spec}, allowing for many successful BBH
simulations across the parameter space
\cite{Boyle:2019kee,Scheel:2025jct,SXSCatalogwebsite}. However, this
scheme relied on Newtonian approximations that prevent it from capturing
important couplings of the controlled parameters. Additionally, it did not allow
the direct control of the system's total energy and angular momentum, which
is useful to parametrize hyperbolic encounters. Moreover, \spec{}'s design is
not optimized for parallelism, which limits performance in large-scale
simulations. \spectre{}~\cite{spectre} is the new task-based NR code developed
by the SXS collaboration designed for better scalability and accuracy, which
will be essential for the next generation of gravitational-wave detectors. As part
of an effort to make \spectre{} capable of fully simulating BBHs
\cite{Lovelace:2025}, we implement in this work an improved control scheme that
builds upon the approach used in \spec{} while addressing its limitations.
We also introduce a way to generate initial data for BBH hyperbolic encounters
by controlling energy and angular momentum.

This paper is organized as follows. In Sec.~\ref{sec:methods}, we describe how
free data are specified before solving the XCTS equations, how physical
parameters are computed from the resulting initial data, and how these
parameters are controlled in both bound and unbound BBH configurations. In
Sec.~\ref{sec:results}, we present numerical results demonstrating the
effectiveness of the control scheme, including a comparison with \spec{}'s
implementation. We summarize and discuss our results in
Sec.~\ref{sec:conclusion}. Finally, we explain how we improve accuracy of our
asymptotic quantities in Appendices \ref{sec:appendix-1} and
\ref{sec:appendix-2}.

\section{Methods}\label{sec:methods}

\subsection{Free data}\label{sec:free_data}

The extended conformal thin-sandwich (XCTS) formulation
(see \cite{Pfeiffer:2004nc,BaumgarteShapiro} for reviews)
decomposes the spatial metric $\gamma_{ij}$ into a conformal factor $\psi$ and
an analytic conformal background metric $\bar\gamma_{ij}$,
\begin{equation}
  \gamma_{ij} = \psi^4 \bar\gamma_{ij}.
\end{equation}
Combined with suitable decompositions of the extrinsic curvature, Einstein's constraint equations become a system of five coupled
elliptic PDEs, given by
\begin{subequations}\label{eq:XCTS}
  \begin{align}\label{eq:Ham}
    \bar\nabla^2 \psi &= \frac{1}{8} \psi \bar R + \frac{1}{12} \psi^5 K^2 \\ \notag
                      &\quad - \frac{1}{8} \psi^{-7} \bar A_{ij} \bar A^{ij} - 2 \pi \psi^5 \rho, \\
    \bar\nabla_i (\bar L \beta)^{ij} &= (\bar L \beta)^{ij} \bar\nabla_i \ln(\bar\alpha) + \bar\alpha \bar\nabla_i (\bar\alpha^{-1} \bar u^{ij}) \\ \notag
                                     &\quad + \frac{4}{3} \bar\alpha \psi^6 \bar\nabla^j K + 16 \pi \bar\alpha \psi^{10} S^j, \\
    \bar\nabla^2(\alpha\psi) &= \alpha\psi \Bigg( \frac{7}{8} \psi^{-8} \bar A_{ij} \bar A^{ij} + \frac{5}{12} \psi^4 K^2 + \frac{1}{8} \bar R \\ \notag
                             &\quad + 2 \pi \psi^4 (\rho + 2 S) \Bigg)  - \psi^5 \partial_t K + \psi^5 \beta^i \bar\nabla_i K,
  \end{align}
\end{subequations}
where $\bar A^{ij} = \frac{1}{2 \bar\alpha} \left( (\bar L \beta)^{ij} - \bar
u^{ij} \right)$ and $\bar \alpha = \alpha \psi^{-6}$. The conformal metric
$\bar\gamma_{ij}$ defines a background geometry in which we define the
covariant derivative $\bar\nabla$, the Ricci scalar $\bar R$,
and the longitudinal operator
\begin{equation}
  (\bar L \beta)^{ij} = \bar\nabla^i \beta^j + \bar\nabla^j \beta^i - \frac{2}{3} \bar\gamma^{ij} \bar\nabla_k \beta^k.
\end{equation}

Equations \eqref{eq:XCTS} must be solved for the conformal factor $\psi$, the lapse
$\alpha$, and the shift $\beta^i$. Prior to that, we must specify the conformal
spatial metric $\bar\gamma_{ij}$, the extrinsic curvature trace $K$, and their
respective time derivatives $\bar u_{ij} \equiv \partial_t \bar\gamma_{ij}$ and
$\partial_t K$. For non-vacuum spacetimes, we must also specify the matter
sources $\rho$, $S$, and $S^i$. It is the fact that $\bar\gamma_{ij}$, $K$,
$\partial_t \bar\gamma_{ij}$, and $\partial_t K$ are freely specifiable
together with boundary conditions that allows us to control the physical
parameters of the initial data.

One particularly successful approach to specify
the free data in the XCTS equations is the superposed Kerr-Schild (SKS)
formulation \cite{Lovelace2008-sw}. It enforces quasiequilibrium conditions by
setting $\partial_t \bar\gamma_{ij}=0$ and $\partial_t K=0$
(see also~\cite{gr-qc/0108076,Cook2004-yf}). Furthermore, it
specifies $\bar\gamma_{ij}$ and $K$ by superposing two analytic solutions of
Kerr-Schild black holes:
\begin{align}
  \bar\gamma_{ij} &= \delta_{ij} + \sum_a e^{-r_a^2/w_a^2} (\gamma_{ij}^{a} - \delta_{ij}), \label{eq:SKS-metric} \\
  K &= \sum_a e^{-r_a^2/w_a^2} K_{a}, \label{eq:SKS-extrinsic-curvature}
\end{align}
where $\delta_{ij}$ is a flat metric, $a \in \{A,B\}$ is used to label each black hole,
$r_a$ is the coordinate distance from the center of black hole $a$, and $w_a$ is a Gaussian decay
parameter. The analytic solutions $\gamma_{ij}^a$ and $K_a$ depend on the mass
parameters $\bar M_a$, dimensionless
spin parameters $\bar\chi^i_a$, and coordinate locations $\vec{c}_a$ of the isolated black holes
We will refer to these parameters as ``conformal'' masses and spins because they are
used to construct the conformal background metric $\bar \gamma_{ij}$.
Empirically, we found that adjusting $\bar M_a$ in the iterative control
procedure, while keeping $\bar\chi^i_a$ fixed to the target spins, is the most
effective way of controlling the physical masses of the black holes.

To avoid singularities in our BBH computational domain, we excise two deformed
spheres representing the black holes, denoted by $S_A$ and $S_B$, which are
centered at coordinate locations $\vec{c}_A+\vec C_0$ and
$\vec{c}_B+\vec C_0$.  Here, $\vec c_A = 1/(q+1)\vec D_0$ and $\vec
c_B = -q/(q+1)\vec D_0$ are the underlying positions of Newtonian
point masses with mass-ratio $q$ and separation
$\vec D_0$, whereas $\vec C_0$ allows to shift the configuration in
response to nonlinear general relativistic effects.
The shape of the excisions is deformed to conform to an ellipsoid of constant
Boyer-Lindquist coordinate $r_a=\bar M_a (1 + \sqrt{1 - \bar{\chi}_a^2})$ \cite{BaumgarteShapiro}.
This means that the control of the conformal masses $\bar M_a$ also adjusts the
size of the excisions. Note that the coordinate center of the excisions is not affected
by this deformation.

As in any elliptic PDE problem, we have to impose boundary conditions on our
variables $\{\psi, \alpha, \beta^i\}$ before solving the XCTS equations. At the
excision surfaces $S_a \in \{S_A, S_B\}$, we impose apparent-horizon
boundary conditions \cite{gr-qc/0108076,Cook2004-yf, Varma2018-fp, Vu:2024}:
\begin{subequations}
  \begin{align}
    \bar n^k \bar \nabla_k \psi \Big|_{S_a} &=
      \frac{\psi^3}{8 \alpha} \bar n_i \bar n_j \left( (\bar L \beta)^{ij} - \bar u^{ij} \right)
      \\ \notag &\qquad
      - \frac{\psi}{4} \bar m^{ij} \bar\nabla_i \bar n_j
      - \frac{1}{6} K \psi^3
      - \frac{\psi^3}{4} \Theta_a,
    \\
    \beta^i \Big|_{S_a} &= \label{eq:shift-inner-bc}
      - \frac{\alpha}{\psi^2} \bar n^i
      + \left( \vec\Omega_a \times (\vec x - \vec c_a) \right)^i
      \\ \notag &\qquad
      + \frac{\bar n^i}{\psi^2} (\bar n_j \beta_a^j + \alpha_a),
    \\
    \alpha \Big|_{S_a} &= \alpha_a
  \end{align}
\end{subequations}
where $\bar n^i$ is the unit normal to the excision surface pointing out of the
computational domain towards the excision center $\vec c_a$ and normalized
with the conformal metric $\bar \gamma_{ij}$,
$\bar m_{ij} = \bar \gamma_{ij} - \bar n_i \bar n_j$ is the induced conformal 2-metric of
$S_a$, $\vec \Omega_a$ is a freely specifiable horizon rotation parameter, and
$\vec x$ are the spatial coordinates. The expansion of the excision surface,
$\Theta_a$, can be set to zero (making the excisions apparent horizons) or to a
negative value (placing the excisions inside apparent horizons). The latter has
been shown to reduce constraint violations during evolution \cite{Varma2018-fp}.
To set a negative expansion we place the excision surface a small fraction
inside $r_a$ and evaluate $\Theta_a$ using the isolated Kerr solution.
The lapse $\alpha_a$ and shift $\beta_a^i$ are also evaluated using the isolated Kerr
solution. We use the horizon
rotations $\vec\Omega_A$ and $\vec\Omega_B$ in Eq.~\eqref{eq:shift-inner-bc}
to control the black hole spins.

The outer boundary of our computational domain, denoted by $S_\infty$, is
placed at a finite radius of $R\sim 10^5M$~\cite{Vu:2024}.
We impose asymptotic flatness at $S_\infty$ using Robin boundary conditions
as detailed in Ref.~\cite{Vu:2024}, so the error incurred by the finite outer
radius is of order $1/R^2$ and therefore below the numerical error of our
simulation.
To control the orbital dynamics of the binary \cite{Pfeiffer:2007yz} we decompose the shift
as $\beta^i = \beta_\mathrm{bg} + \beta^i_\mathrm{excess}$, impose the outer boundary
conditions on $\beta^i_\mathrm{excess}$, and set the background shift to
\begin{equation} \label{eq:shift-outer-bc}
  \beta^i_\mathrm{bg} = (\vec \Omega_0 \times \vec x)^i + \dot a_0 x^i + v_0^i
\end{equation}
throughout the domain.
We use the orbital angular velocity $\vec\Omega_0$,
the radial expansion velocity $\dot a_0$, and the linear velocity $\vec v_0$ to
control the initial kinematics (energy and momenta) of the system.
Note that the first term in Eq.~(\ref{eq:shift-outer-bc}) implies a rotation
about the origin of the coordinate system, which ties with the use of the
offset $\vec C_0$ to control the center of mass to be the origin.

One might worry about potential overall rotations of the coordinate system if we
let $\vec\Omega_0$ and $\vec D_0$ have arbitrary directions.
We fix this by placing the black holes near the $xy$-plane, forcing their
initial motion to be parallel to the $xy$-plane with $\vec\Omega_0 = (0,0,\Omega_0^z)$,
and setting their initial separation vector to be parallel to the $x$-axis with
$\vec D_0 = (D_0,0,0)$.
This reduces the six degrees of freedom in $\vec\Omega_0$ and $\vec D_0$ to
only two.

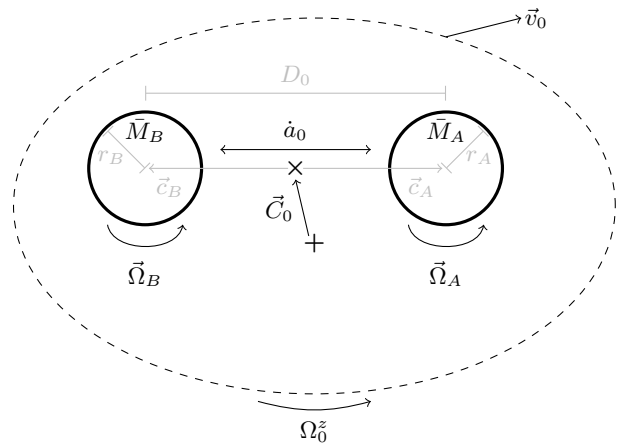
\begin{figure}
  \centering
  \begin{tikzpicture}
    \draw[very thick] (-2,0) circle (0.75cm);
    \draw[very thick] (+2,0) circle (0.75cm);

    \node at (-2,0.5) {$\bar M_B$};
    \node at (+2,0.5) {$\bar M_A$};

    \node at (-2,-1.4) {$\vec\Omega_B$};
    \node at (+2,-1.4) {$\vec\Omega_A$};
    \draw[->] (-2.5,-0.75) to [out=-80,in=-100] (-1.5,-0.75);
    \draw[->] (1.5,-0.75) to [out=-80,in=-100] (2.5,-0.75);
    
    \draw[->,lightgray] (-0.1,0) -- (-1.95,0) node[pos = 0.85, anchor = north] {$\vec c_B$};
    \draw[->,lightgray] (+0.1,0) -- (+1.95,0) node[pos = 0.85, anchor = north] {$\vec c_A$};
    \draw[|-|,lightgray] (-2,1) -- (+2,1) node[pos = 0.5, anchor = south] {$D_0$};

    \draw[|-|,lightgray] (-2,0) -- (-2.5,0.5) node[pos = 0.3, anchor = east] {$r_B$};
    \draw[|-|,lightgray] (+2,0) -- (+2.5,0.5) node[pos = 0.3, anchor = west] {$r_A$};

    \node at (0,0) {$\boldsymbol\times$};
    \node at (0.25,-1) {$\boldsymbol+$};
    \draw[->] (0.175,-0.9) -- (0,-0.15) node[pos = 0.5, anchor = east] {$\vec C_0$};

    \node at (0,0.5) {$\dot a_0$};
    \draw[<->] (-1,0.25) -- (1,0.25);

    \draw[dashed] (0.25,-0.5) ellipse [x radius=4cm, y radius=2.5cm];
    
    \node at (3.2,2) {$\vec v_0$};
    \draw[->] (2,1.75) -- (3,2);

    \node at (0.25,-3.5) {$\Omega_0^z$};
    \draw[->] (-0.5,-3.1) to [out=-15,in=-165] (1,-3.1);
  \end{tikzpicture}
  \caption{
    Schematic representation of the BBH free data.
    The solid circles represent the two black hole excisions ($S_A$ and $S_B$),
    while the dashed ellipse represents the outer boundary of the computational domain ($S_\infty$).
    ``$\boldsymbol+$'' and ``$\boldsymbol\times$'' indicate the origin and the
    Newtonian center of mass, respectively.
    The gray free data are not explicitly used in the \spectre{} control scheme.
  }
  \label{fig:free-data}
\end{figure}

These free data choices are summarized schematically in
Fig.~\ref{fig:free-data}. Note that other sets of free data can be used to
control the same physical parameters. For example, \spec{} uses the excision
radii $r_a$ and their centers $\vec c_a$ to control the black hole masses and their positions, respectively \cite{Ossokine:2015yla}. We have also tried to use
$\bar\chi^i_a$ to control the black hole spins instead of
$\vec\Omega_a$, but we found that the latter allowed us to achieve higher
spins and mass ratios.

\subsection{Physical parameters}

Once the XCTS system in Eqs.~\eqref{eq:XCTS} is solved, we can measure physical
parameters of the initial data. We use a fast flow method for finding apparent
horizons based on Ref.~\cite{Gundlach1997us}. Using the resulting horizons, we can
compute their Christodoulou masses $M_{A,B}$ \cite{Christodoulou:1970} and
dimensionless spins $\vec\chi_{A,B}$ \cite{Owen:2019}.

Famously, global quantities like energy and momentum are nontrivial to define in
general relativity. Fortunately, we can use the Arnowitt-Deser-Misner (ADM)
formalism (see \cite{BaumgarteShapiro} for a review) to define total energy
$E_\text{ADM}$, linear momentum $P_\text{ADM}^i$, and angular momentum
$J^\text{ADM}_i$ for asymptotically flat spacetimes. These definitions take the
form of surface integrals evaluated at spatial infinity and are given by
\cite[Eqs.~3.131--3.195]{BaumgarteShapiro}
\begin{align}
  E_\text{ADM} &= \frac{1}{16\pi} \oint_{S_\infty}  \Big(
                    \gamma^{jk} \Gamma^i_{jk}
                    - \gamma^{ij} \Gamma^k_{jk}
                  \Big) \, dS_i, \label{eq:ADM-energy} \\
  P_\text{ADM}^i &= \frac{1}{8\pi} \oint_{S_\infty} \Big( K^{ij} - K \gamma^{ij} \Big) \, dS_j, \label{eq:ADM-linear-momentum} \\
  J^\text{ADM}_i &= \frac{1}{8\pi} \oint_{S_\infty} \epsilon_{ijk} x^j \Big( K^{kl} - K \gamma^{kl} \Big) \, dS_l, \label{eq:ADM-angular-momentum}
\end{align}
where $\Gamma^i_{jk}$ are the Christoffel symbols associated with $\gamma_{ij}$.

Since we compute these asymptotic quantities on a large-radius surface at
the outer boundary placed at $r\sim 10^5$, the results are sensitive to
numerical errors from large area elements.
Additionally, we can express the integrands in terms of the analytic quantities
used in the conformal background, avoiding some numerical derivatives and
improving accuracy.
We reformulate Eqs.~\eqref{eq:ADM-energy}--\eqref{eq:ADM-angular-momentum}
in Appendix~\ref{sec:appendix-1} to address these issues.

Using the formalism developed by Baskaran et al.~\cite{Baskaran:2003}
and assuming conformal flatness, we define an asymptotic center of mass
given by
\begin{equation}\label{eq:center-of-mass}
  C_\text{CoM}^i = \frac{3}{2 \pi E_\text{ADM}}
                     \oint_{S_\infty} (\psi-1) \tilde n^i \, d\tilde A,
\end{equation}
where $\tilde n^i = x^i / r$ is the Euclidean outward-pointing unit normal
and $d\tilde A$ is the Euclidean area element of $S_\infty$.
One way to interpret Eq.~\eqref{eq:center-of-mass} is that we are summing over
the unit vectors $\tilde n^i$, rescaled by $\psi$, in all directions.
If $\psi$ is constant everywhere, no rescaling happens and all the unit vectors
cancel out. If $\psi$ is larger in some region (e.g., near a black hole), then
the vectors in this region dominate, giving off the center of mass as a result.
We show the derivation of Eq.~\eqref{eq:center-of-mass} from
Ref.~\cite{Baskaran:2003} in Appendix~\ref{sec:appendix-2}.

\subsection{Control of bound orbits}\label{sec:control-bound}

To construct initial data for a bound-orbit inspiral BBH system, we begin by
specifying target masses $M_A^*$ and $M_B^*$ and target dimensionless spins
$\vec\chi_A^*$ and $\vec\chi_B^*$ for each black hole. Additionally, we want to
eliminate motions of the system by driving $\vec C_\text{CoM}$ and $\vec
P_\text{ADM}$ to zero. This minimizes drifts in the binary's orbit, especially
for long simulations.

Let the choice of free data be represented as
\begin{equation}
  {\bf u} = \Big( \bar M_A, \bar M_B, \vec\Omega_A, \vec\Omega_B, \vec C_0, \vec v_0 \Big).
\end{equation}
Also, let the difference between the measured and target physical parameters be
represented by the residual function
\begin{equation}
  \begin{aligned}
    {\bf F}({\bf u}) = \Big( M_A - M^*_A, M_B - M^*_B, \vec\chi_A - \vec\chi^*_A, \quad\\ \vec\chi_B - \vec\chi^*_B, \vec C_\text{CoM}, \vec P_\text{ADM} \Big).
  \end{aligned}
\end{equation}
Note that there are 14 components in ${\bf u}$ and ${\bf F}$.
We order elements such that a component of ${\bf u}$ primarily affects the
corresponding component of ${\bf F}$.
The choice of $\dot a_0$, $\Omega_0^z$, and $D_0$ parametrizes the orbit, which
gets controlled separately via eccentricity control~\cite{Knapp:2024yww,Nee:2025zdy}
or eccentricity reduction~\cite{Habib:2025} schemes.

A natural choice of initial guesses for the free data is to use the target
values. For a single Kerr black hole (labeled by $a \in \{A,B\}$), we know that
\begin{align}
  \vec\chi_a &= - 2 r_a \vec\Omega_a, \label{eq:Kerr-spin} \\
  r_a &= M_a \left(1 + \sqrt{1 - |\vec\chi_a|^2}\right) \label{eq:Kerr-radius},
\end{align}
where $r_a$ is the outer horizon radius in Boyer-Lindquist radial coordinates.
Then, we can define an initial horizon rotation as
\begin{equation}
  \vec\Omega^*_a = - \frac{\vec\chi^*_a}{2 M^*_a \left(1 + \sqrt{1 - |\vec\chi^*_a|^2}\right)}.
\end{equation}
This gives us the initial guess
\begin{equation}
  {\bf u}_0 = \Big( M^*_A, M^*_B, \vec\Omega^*_A, \vec\Omega^*_B, \vec 0, \vec 0 \Big).
\end{equation}

To drive ${\bf F}({\bf u})$ to zero, we can update our free data at every
control iteration $k$ using a Newton-Raphson scheme~\cite{NumericalRecipes}:
\begin{equation}\label{eq:NewtonRaphson}
  {\bf u}_{k+1} = {\bf u}_{k} - \mathbb{J}^{-1}_{k} \cdot {\bf F}_{k},\quad k\ge 0.
\end{equation}
Each evaluation of ${\bf F}({\bf u})$ is computationally
expensive because it requires an entire initial data solve.  Therefore, computing
the Jacobian in Eq.~\eqref{eq:NewtonRaphson} is unfeasible. That
said, we can iteratively find an approximation $\mathbb{J}_k$ from an initial guess using
Broyden's method~\cite{NumericalRecipes}:
\begin{equation}\label{eq:Broyden}
  \mathbb{J}_k = \mathbb{J}_{k-1} + \frac{{\bf F}_k \otimes \Delta {\bf u}_k}{||\Delta {\bf u}_k||^2}, \quad k\ge 1,
\end{equation}
where $\Delta {\bf u}_k = {\bf u}_k - {\bf u}_{k-1}$, and $\otimes$ indicates
an outer product.

Since we approximate the Jacobian iteratively using Eq.~\eqref{eq:Broyden}, we must
choose an initial guess $\mathbb{J}_0$. To motivate our derivation, let us assume
that the measured masses are close to the conformal masses (i.e., $M_a \approx
\bar M_a$), which implies that the diagonal element of the Jacobian in
the $M_a$ dimension is given by
\begin{equation}\label{eq:J-mass-approx}
  \frac{\partial (M_a - M^*_a)}{\partial \bar M_a}\Bigg|_{k=0} \approx 1.
\end{equation}
Differentiating the Kerr expressions in
Eqs.~\eqref{eq:Kerr-spin}--\eqref{eq:Kerr-radius}, we obtain
\begin{align}
  \frac{\partial (\chi^i_a - \chi^{*,i}_a)}{\partial \bar M_a}\Bigg|_{k=0} &\approx -2 \Omega^{*,i}_a \left(1 + \sqrt{1 - |\vec\chi^*_a|^2}\right), \label{eq:J-chi-mass} \\
  \frac{\partial (\chi^i_a - \chi^{*,i}_a)}{\partial \Omega^j_a}\Bigg|_{k=0} &\approx -2 \bar M_a \left(1 + \sqrt{1 - |\vec\chi^*_a|^2}\right) \delta^i_j.
\end{align}
Under a Newtonian approximation for the center of mass, we have
\begin{equation} \label{eq:CoM-Newtonian}
  \vec C_\text{CoM} \approx \frac{M_A \vec c_A + M_B \vec c_B}{M_A + M_B} + \vec C_0,
\end{equation}
where we choose $\vec c_A$ and $\vec c_B$ so that the first term in
Eq.~\eqref{eq:CoM-Newtonian} is zero assuming $M_a \approx M^*_a$.
This leads to
\begin{equation}
  \frac{\partial C_\text{CoM}^i}{\partial C_0^j}\Bigg|_{k=0} \approx \delta^i_j.
\end{equation}
Similarly, assuming the boundary condition \eqref{eq:shift-outer-bc}
sets the center-of-mass velocity, we have
\begin{equation}\label{eq:Padm-Newtonian}
  \vec P_\text{ADM} \approx (M_A + M_B) \Big[ \vec \Omega_0 \times \vec C_\text{CoM} + \dot a_0 \vec C_\text{CoM} + \vec v_0 \Big].
\end{equation}
Note that we enforce $M^*_A + M^*_B = 1$, so we can assume
$M_A + M_B \approx 1$ when initializing $\mathbb{J}_0$. Then,
\begin{equation}\label{eq:J-Padm-approx}
  \frac{\partial P_\text{ADM}^i}{\partial v_0^j}\Bigg|_{k=0} \approx \delta^i_j.
\end{equation}
We disregard other cross-terms at $k=0$ and let Broyden's method adjust the
remaining Jacobian terms as needed.

\subsection{Control of hyperbolic encounters}\label{sec:control-hyperbolic}

For hyperbolic BBH encounters it is convenient to parametrize the system in
terms of global, asymptotic quantities, which allows for unambiguous comparisons
with other NR codes and perturbative calculations 
\cite{Rettegno:2023ghr,Buonanno:2024byg,Swain:2024ngs,Long:2025nmj}.
Specifically, we drive $E_\text{ADM}$ and $J^\text{ADM}_z$ to their target
values $E^*$ and $J^*$.
We focus only on the $z$-component of $\vec J^\text{ADM}$ because we target
orbits initially in the $xy$-plane. The control scheme presented here is the
first direct control of ADM energy and angular momentum in NR simulations. It was
initially implemented in \spec{} (used for the study of BBH scattering angles in
Ref.~\cite{Long:2025nmj}) and later adapted to \spectre{}.

In addition to controlling the energy and angular momentum of the system, we
still want to control the same quantities as in the bound case. Then, let the
choice of free data be represented as
\begin{equation}
  {\bf u} = \Big( \bar M_A, \bar M_B, \vec\Omega_A, \vec\Omega_B, \vec C_0, \vec v_0, \dot a_0, \Omega_0^z \Big).
\end{equation}
Similarly, let the residual function be
\begin{equation}
  \begin{aligned}
    {\bf F}({\bf u}) = \Big( M_A - M^*_A, M_B - M^*_B, \vec\chi_A - \vec\chi^*_A, \vec\chi_B - \vec\chi^*_B, \quad\\ \vec C_\text{CoM}, \vec P_\text{ADM}, E_\text{ADM} - E^*, J^\text{ADM}_z - J^* \Big).
  \end{aligned}
\end{equation}
Note that there are now 16 components in ${\bf u}$ and ${\bf F}$.

We use the same iterative procedure from
Eqs.~\eqref{eq:NewtonRaphson}--\eqref{eq:Broyden}. To choose an initial guess
for $\dot a_0$ and $\Omega^z_0$, we use Newtonian approximations for the total
energy and angular momentum:
\begin{align}
  E_\text{ADM} &\approx M + \frac{1}{2} \mu \dot a_0^2 D_0^2, \label{eq:Eadm-Newtonian} \\
  J^\text{ADM}_z &\approx \mu D_0^2 \Omega_0^z, \label{eq:Jadm-Newtonian}
\end{align}
where $M = M_A + M_B$ is the total horizon mass, $\mu=M_AM_B/(M_A+M_B)$ is the
reduced mass.
In Eqs.~\eqref{eq:Eadm-Newtonian}--\eqref{eq:Jadm-Newtonian}, we are assuming
large separation and that the hyperbolic incoming motion is nearly radial, which 
results in $E_\text{ADM}$ and $J^\text{ADM}_z$ being dominated by $\dot a_0$
and $\Omega_0^z$, respectively.
With this, we initialize the free data as
\begin{equation}
  {\bf u}_0 = \Big( M^*_A, M^*_B, \vec\Omega^*_A, \vec\Omega^*_B, \vec 0, \vec 0, \dot a_0^*, \Omega_0^{z,*} \Big),
\end{equation}
where
\begin{align}
  \dot a_0^* &= - \sqrt{2 \frac{E^* - M}{\mu D_0^2}}, \\
  \Omega_0^{z,*} &= \frac{J^*}{\mu D_0^2}.
\end{align}
The sign of $\dot a_0^*$ is negative such that the binary is initially
on the ingoing leg of the hyperbola.

For the initialization of $\mathbb{J}_0$, we assume -- as in the bound case --
that $M_{A,B} \approx \bar M_{A,B}$, so the
approximations in Eqs.~\eqref{eq:J-mass-approx}--\eqref{eq:J-Padm-approx} still
apply. We can differentiate the Newtonian approximations in
Eqs.~\eqref{eq:Eadm-Newtonian}--\eqref{eq:Jadm-Newtonian} to get the remaining
diagonal elements of $\mathbb{J}_0$:
\begin{align}
  \frac{\partial E_\text{ADM}}{\partial \dot a_0} &\approx \mu \dot a_0 D_0^2, \label{eq:J-Eadm-adot0} \\
  \frac{\partial J^\text{ADM}_z}{\partial \Omega_0^z} &\approx \mu D_0^2.
\end{align}
Additionally, we obtain some off-diagonal elements of $\mathbb{J}_0$ by differentiating
Eqs.~\eqref{eq:Eadm-Newtonian}--\eqref{eq:Jadm-Newtonian} relative to the
horizon masses:
\begin{align}
  \frac{\partial E_\text{ADM}}{\partial \bar M_A} &\approx 1 + \frac{1}{2} \frac{M_B^2}{M^2} \dot a_0^2 D_0^2, \label{eq:J-Eadm-mass-a} \\
  \frac{\partial E_\text{ADM}}{\partial \bar M_B} &\approx 1 + \frac{1}{2} \frac{M_A^2}{M^2} \dot a_0^2 D_0^2, \\
  \frac{\partial J^\text{ADM}_z}{\partial \bar M_A} &\approx \frac{M_B^2}{M^2} D_0^2 \Omega_0^z, \\
  \frac{\partial J^\text{ADM}_z}{\partial \bar M_B} &\approx \frac{M_A^2}{M^2} D_0^2 \Omega_0^z, \label{eq:J-Jadm-mass-b}.
\end{align}
Note that we can use $M \approx 1$, $\dot a_0 \approx \dot a_0^*$, and
$\Omega_0^z \approx \Omega_0^{z,*}$ in
Eqs.~\eqref{eq:J-Eadm-adot0}--\eqref{eq:J-Jadm-mass-b}.

This approach permits -- and predicts -- the coupling between the ADM energy and
angular momentum with the horizon masses via off-diagonal terms in the Jacobian.
We use the Newtonian approximations in
Eqs.~\eqref{eq:J-Eadm-mass-a}--\eqref{eq:J-Jadm-mass-b} as initial estimates
for these couplings and allow all Jacobian terms to develop through
Broyden's method.

\subsection{Comparison with \spec{}}\label{sec:comparison-spec}

In contrast with the parameter control scheme presented in
Secs.~\ref{sec:control-bound}--\ref{sec:control-hyperbolic}, \spec{} only
applies Broyden's method for horizon quantities (i.e., masses and spins).
To control asymptotic quantities, it expands Newtonian approximations under
small perturbations, leading to fixed Newtonian updating formulas.

\spec{}'s current control of bound orbits was introduced in
Ref.~\cite{Ossokine:2015yla}, improving on Refs.~\cite{Buonanno:2010yk,Buchman2012-ud}.
Considering perturbations of $\vec c_A$ and $\vec v_0$,
Eq.~\eqref{eq:CoM-Newtonian} and Eq.~\eqref{eq:Padm-Newtonian}
lead to \cite{Ossokine:2015yla}
\begin{align}
  \vec c_{A,k+1} &= \label{eq:SpEC-CoM-control}
    \vec c_{A,k}
    - \vec C_{\text{CoM}, k}
    \\ \notag &\quad
    - \frac{M_{A,k} \Delta M_{B,k} - M_{B,k} \Delta M_{A,k}}{(M_{A,k} + M_{B,k})^2} \vec D_0,
  \\
  \vec v_{0,k+1} &= \label{eq:SpEC-Padm-control}
    \vec v_{0,k}
    - \frac{\vec P_{\text{ADM},k}}{(M_{A,k} + M_{B,k})}
    \\ \notag &\quad
    + (\Delta M_{A,k} + \Delta M_{B,k}) (\vec v_{0,k} + \vec \Omega_{0,k} \times \vec c_{A,k})
    \\ \notag &\quad
    - \vec \Omega_{0,k} \times \Delta \vec c_{A,k}
    - \frac{\Delta M_{B,k}}{(M_{A,k} + M_{B,k})} \vec \Omega_{0,k} \times \vec D_0.
\end{align}
Note that \spec{} chooses to use $\vec c_A$ instead of $\vec C_0$ to control
$\vec C_\text{CoM}$, setting $\vec c_B=\vec c_A-\vec D_0$.

In order to permit the study of BBH scattering angles \cite{Long:2025nmj},
a control of hyperbolic encounters was implemented in \spec{} following similar
ideas from Ref.~\cite{Ossokine:2015yla}.
Expanding Eqs.~\eqref{eq:Eadm-Newtonian}--\eqref{eq:Jadm-Newtonian} under
perturbations of $\dot a_0$ and $\Omega_0^z$, we have
\begin{align}
  \dot a_{0,k+1} &= \label{eq:SpEC-Eadm-control}
    \dot a_{0,k}
    + \frac{E^* - E_k}{\mu \dot a_{0,k} D_0^2},
  \\
  \Omega^z_{0,k+1} &= \label{eq:SpEC-Jadm-control}
    \Omega^z_{0,k}
    + \frac{J^* - J_k}{\mu D_0^2},
\end{align}
where $E_k$ and $J_k$ are estimates for the ADM energy and angular momentum after the current updating step, which are introduced to reduce over-adjustments. They are computed as
\begin{align}
  E_k &= E_{\text{ADM},k} - \lambda \sum_a (M_{a,k} - M^*_a), \label{eq:SpEC-projected-E} \\
  J_k &= J^\text{ADM}_{z,k} - \lambda \sum_a (M_{a,k} - M^*_a) ({c_a^x}^2 + {c_a^y}^2) \Omega^z_{0,k}, \label{eq:SpEC-projected-J}
\end{align}
where $a \in \{A,B\}$, and $\lambda=0.5$ is a relaxation factor.

\section{Results}\label{sec:results}

\begin{figure}
  \centering
  \includegraphics[width=\columnwidth]{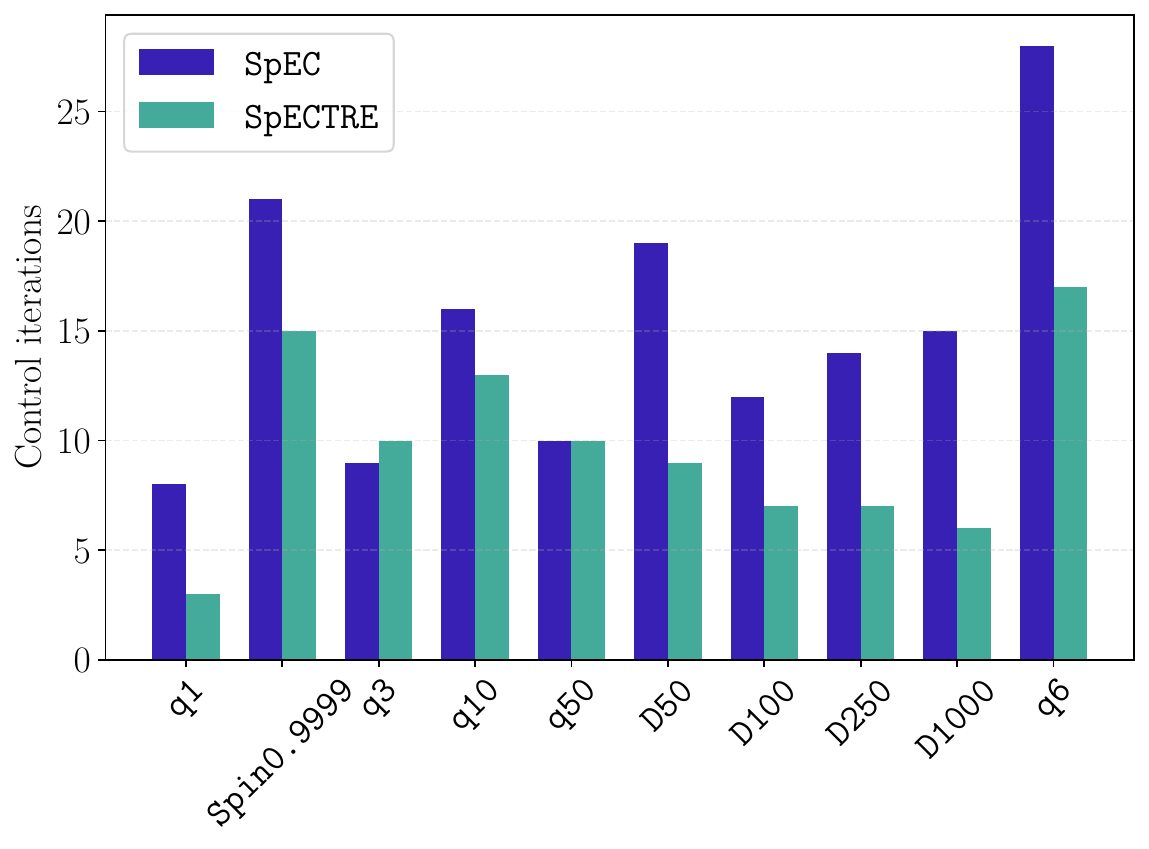}
  \caption{
    \label{fig:iterations-histogram}
    Histogram of the number of control iterations needed for all configurations in Tables \ref{tab:bound-orbits} and \ref{tab:hyperbolic-encounters}.
  }
\end{figure}

We now demonstrate the effectiveness of our control scheme for a series of
both bound and unbound BBH configurations. Table \ref{tab:bound-orbits}
lists bound orbit cases, chosen to match the parameters in
Ref.~\cite{Ossokine:2015yla} (plus \texttt{q1}), while Table \ref{tab:hyperbolic-encounters} lists
hyperbolic encounter cases with $E^*$ and $J^*$ selected to coincide
with simulations from Ref.~\cite{Long:2025nmj}.
All results from \spectre{} use the control scheme
described in Sec.~\ref{sec:control-bound}--\ref{sec:control-hyperbolic}, while
results from \spec{} use the procedure described in Sec.~\ref{sec:comparison-spec}.
Figure \ref{fig:iterations-histogram} summarizes how
many control iterations are needed for all runs in both codes.

\begin{table*}
  \centering
  \begin{tabular}{c|ccc|ccc}
    Name             & $\;q = M^*_A/M^*_B\;$  & $\vec\chi^*_A$          & $\vec\chi^*_B$         & $D_0$   & $\Omega^z_0$ & $\dot a_0$              \\ \hline
    {\tt q1}         & $1$  & $(0,0,0)$               & $(0,0,0)$              & $\;15.00$ & $\;0.01442\;$    & $-4.075 \times 10^{-5}$ \\
    {\tt Spin0.9999} & $1$  & $(0,0,0.9999)$          & $(0,0,0.9999)$         & $14.17$ & $0.01682$    & $-5.285 \times 10^{-5}$ \\
    {\tt q3}         & $3$  & $(0,0.49,-0.755)$       & $(0,0,0)$              & $15.48$ & $0.01515$    & $-3.954 \times 10^{-5}$ \\
    {\tt q10}        & $10$ & $(0.815,-0.203,0.525)$  & $\;(-0.087,0.619,0.647)\;$ & $15.09$ & $0.01542$    & $-1.558 \times 10^{-5}$ \\
    {\tt q50}        & $50$ & $\;(-0.045,0.646,-0.695)\;$ & $(0,0,0)$              & $16.00$ & $0.01428$    & $-3.702 \times 10^{-6}$ \\
  \end{tabular}
  \caption{
    \label{tab:bound-orbits}
    Bound orbit configurations.
  }
\end{table*}

\begin{table*}
  \centering
  \begin{tabular}{c|ccc|ccc}
    Name        & $\;q = M^*_A/M^*_B\;$ & $\vec\chi^*_A$      & $\vec\chi^*_B$ & $D_0$        & $E^*$    & $J^*$         \\ \hline
    {\tt D50}   & $1$                   & $(0,0,0)$           & $(0,0,0)$      & $\;\;50.0$   & $1.0226$ & $1.6039$      \\
    {\tt D100}  & $1$                   & $(0,0,0)$           & $(0,0,0)$      & $100.0$      & $1.0226$ & $1.0941$      \\
    {\tt D250}  & $1$                   & $(0,0,0)$           & $(0,0,0)$      & $\;250.0\;$  & $1.0550$ & $1.2943$      \\
    {\tt D1000} & $1$                   & $(0,0,0)$           & $(0,0,0)$      & $\;1000.0\;$ & $1.0550$ & $1.2943$      \\
    {\tt q6}    & $6$                   & $\;(0.3,0.3,0.3)\;$ & $(0,0,0)$      & $100.0$      & $1.0123$ & $\;\;0.83265$ \\
  \end{tabular}
  \caption{
    \label{tab:hyperbolic-encounters}
    Hyperbolic encounter configurations.
  }
\end{table*}

\subsection{Convergence of physical parameters}

\begin{figure}
  \centering

  \includegraphics[width=\columnwidth]{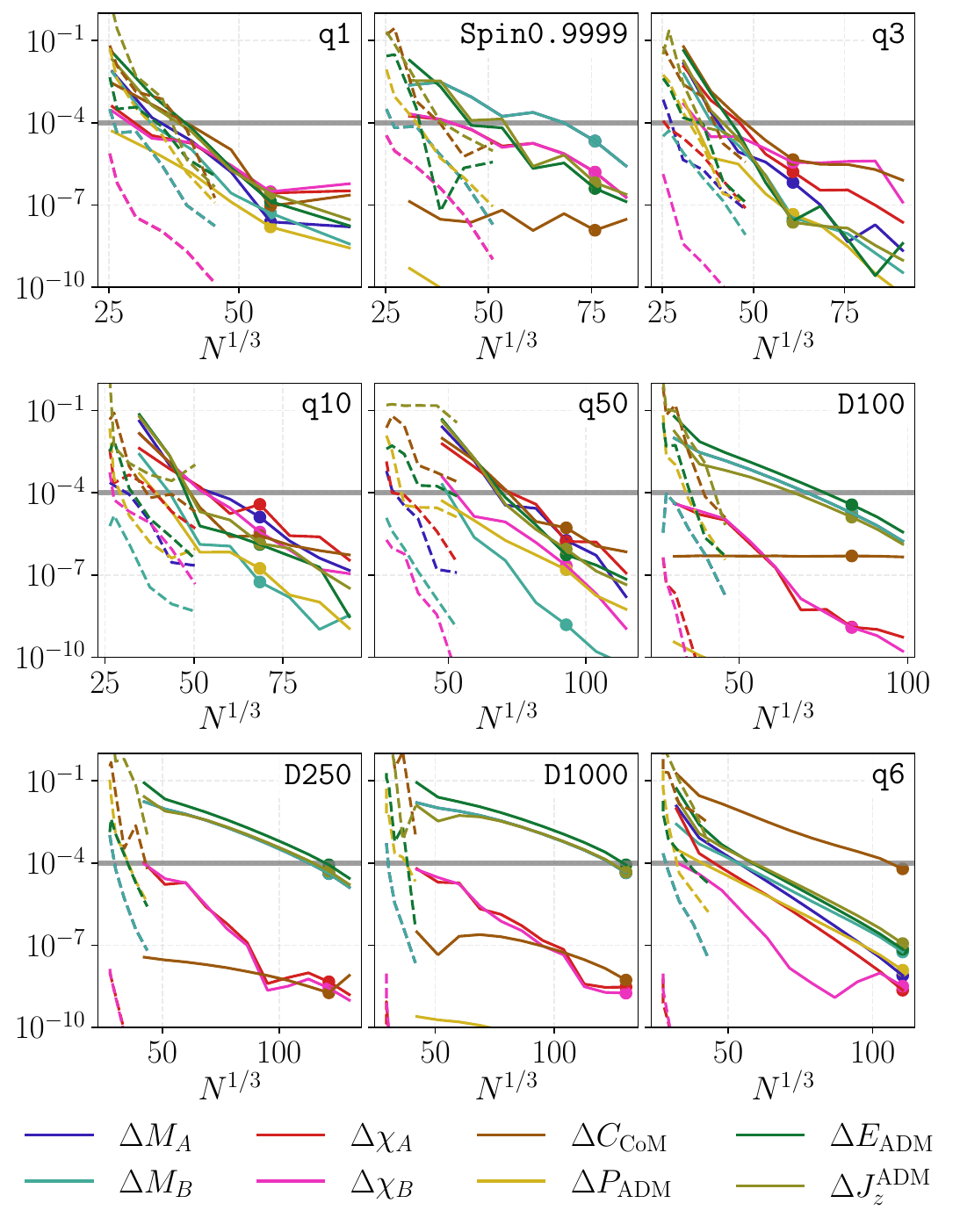}

  \caption{
    \label{fig:resolution-convergence}
    Resolution convergence of physical parameters for the runs in Tables \ref{tab:bound-orbits} and \ref{tab:hyperbolic-encounters}.
    \spectre{} data are shown in solid lines, while \spec{} data are shown in dashed lines.
    The circle marks indicate the resolution chosen for the runs in Figs.~\ref{fig:control-bound}--\ref{fig:control-hyperbolic}.
    The difference of any parameter $x$ with its highest resolution value is
    represented as $\Delta x = x - x|_{\max N}$.
  }
\end{figure}

Before analyzing specific configurations, we verify that the measurements of
the physical parameters converge with resolution. In both bound and hyperbolic
cases, we check the convergence of all parameters as a function of the total
number of grid points $N$ in the computational domain, as shown in
Fig.~\ref{fig:resolution-convergence}.

These convergence tests are used to determine which resolution is required for
each configuration. This is particularly important for \spectre{} because it
currently uses a fixed resolution for the entire control loop, while \spec{}
uses adaptive mesh refinement (AMR) to increase resolution as needed
\cite{Ossokine:2015yla}. We indicate with a circle the resolution chosen in
\spectre{} for the control loops shown later.

In theory, we could control the physical parameters to arbitrary accuracy.
However, in practice, we see that spurious gravitational radiation (``junk
radiation'') produced during the early stages of BBH evolution has an effect on
these parameters on the order of $\sim 10^{-4}$ \cite{Ossokine:2015yla}.
Accordingly, we set our error tolerance to $10^{-4}$ in this work, indicated as
a gray horizontal line in the figures.

From Fig.~\ref{fig:resolution-convergence}, it is clear that \spectre{}
consistently requires more grid points in order to achieve the same accuracy
levels as \spec{}. This is expected because \spectre{} uses a discontinuous
Galerkin method that splits the computational domain into more and smaller
spectral elements than \spec{} in order to achieve better
parallelism~\cite{Vu:2022}.

\subsection{Bound orbits}

To facilitate the comparison between \spectre{} and \spec{}, the BBH
configurations in Table \ref{tab:bound-orbits} were chosen to match the
parameters used in Ref.~\cite{Ossokine:2015yla}, except for the
equal-mass non-spinning case ({\tt q1}) added here for reference. Note that
these cases test extreme regions of the parameter space. In particular, the
{\tt Spin0.9999} and {\tt q50} test cases are beyond the current capability of
\spec{} and \spectre{} to evolve, but potential targets for the future.
Also note that all spinning cases include spin magnitudes larger than $0.9$.

All configurations in Table \ref{tab:bound-orbits} were controlled up to a
residual tolerance of $10^{-4}$ using the latest versions of \spec{} and
\spectre{} \cite{spectre}.
Input files for the results shown in this work are available in
the supplementary material.
Figure~\ref{fig:control-bound} compares the behavior of the control loop in each
code.

\begin{figure}
  \centering
  \includegraphics[width=\columnwidth]{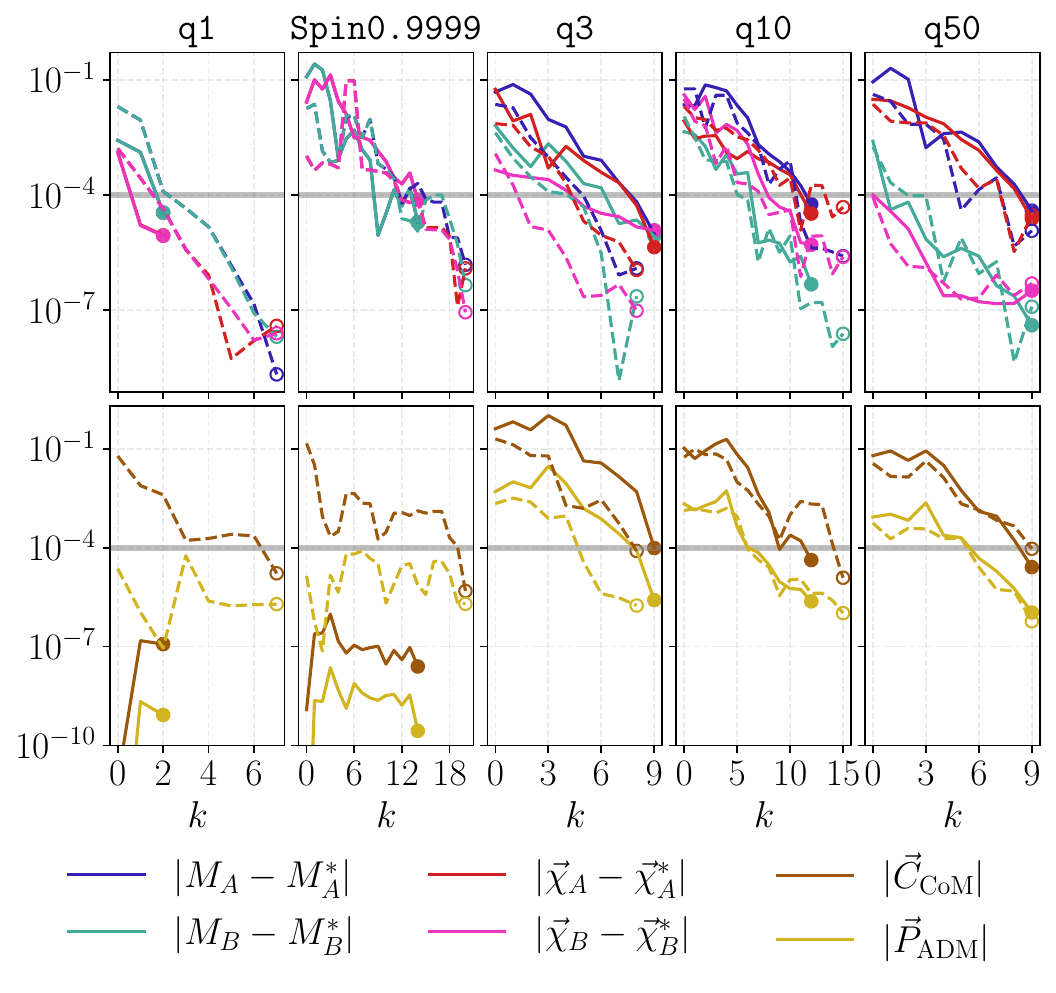}
  \caption{
    \label{fig:control-bound}
    Comparison of control loops in \spec{} (dashed) and \spectre{} (solid) for bound orbits.
    The top panels show results for black hole masses and spins, whereas the bottom panels show the asymptotic quantities.
  }
\end{figure}

For the equal-mass configurations {\tt q1} and {\tt Spin0.9999}, it is clear
that \spectre{} converges significantly faster than \spec{}. However,
this does not necessarily indicate an improvement in the control scheme. In
these cases, $\vec C_\text{CoM}$ is trivially zero due to symmetry. Since
\spec{} starts at low resolution and gradually increases it through AMR, its
truncation error can initially mask this trivial parameter. This is supported by
the fact that the residuals of the measures masses and spins in both codes reach
the tolerance level near the same iteration, after which \spec{} only waits for
$\vec C_\text{CoM}$ to decrease.
In the other cases ({\tt q3}, {\tt q10}, and {\tt q50}), where $\vec
C_\text{CoM}$ is not zero initially, the number of iterations is similar between
the codes, differing at most by 3. This confirms that \spectre{}'s control
scheme successfully reproduces the results of the previous implementation in \spec{}.

\begin{figure}
  \centering
  \includegraphics[width=\columnwidth]{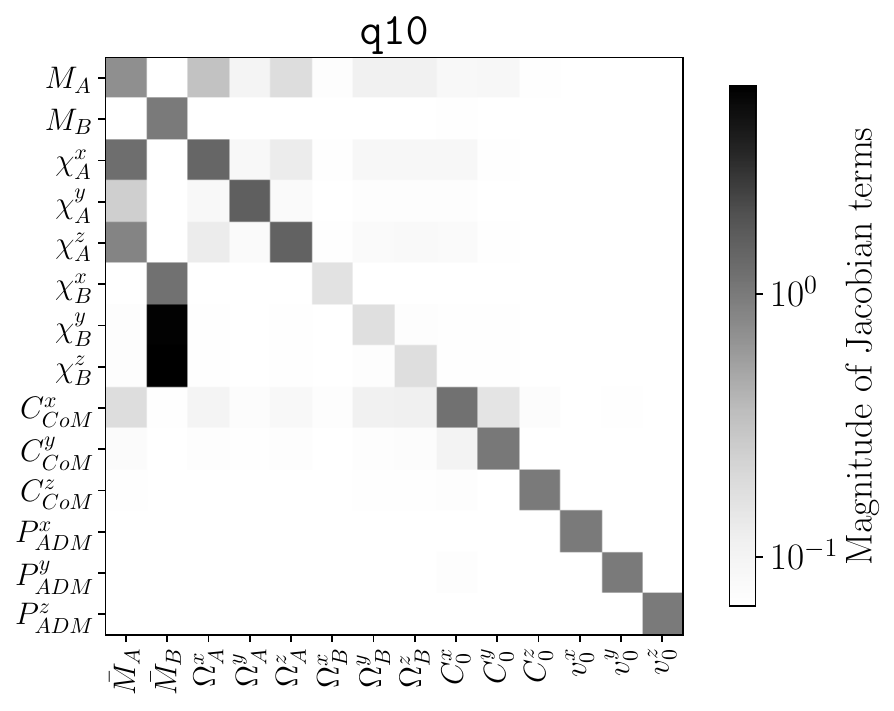}
  \caption{
    \label{fig:jacobian_bound}
    Final control Jacobian of the {\tt q10} case.
  }
\end{figure}

Overall, we see that the Newtonian perturbations done in \spec{} (Sec.~\ref{sec:comparison-spec})
have similar results to the approach taken in \spectre{} (Sec.~\ref{sec:control-bound})
for the control of bound orbits.
To understand why this is the case, we can look at the control Jacobian after
performing the Broyden updates in Eq.~\eqref{eq:Broyden}.
Figure \ref{fig:jacobian_bound} shows the final Jacobian for the {\tt q10} case.
Note that very few off-diagonal terms emerge for $\vec C_\text{CoM}$ and
$\vec P_\text{ADM}$, which explains why the Newtonian expressions in
Eqs.~\eqref{eq:SpEC-CoM-control}--\eqref{eq:SpEC-Padm-control}
are sufficiently accurate approximations.

Figure \ref{fig:jacobian_bound} is also useful for assessing whether our choices
of free data and the initial Jacobian guess were appropriate.
We observe that the Jacobian remains mostly diagonal, indicating that
our choice of free data in Sec.~\ref{sec:free_data} is effective for controlling
these parameters.
Taking the difference between the Jacobian terms in Fig.~\ref{fig:jacobian_bound}
and their initial guesses, we find that the Broyden adjustments are at most on
the order of $\sim 10^{-1}$ and primarily in parameters associated with the large black hole.
It is also clear that the off-diagonal terms in Eq.~\eqref{eq:J-chi-mass}
correctly capture the coupling between the black hole spins and their conformal
masses, which becomes particularly important in high-spin scenarios such as this
one (with spin magnitude $\sim 0.99$).

\subsection{Hyperbolic encounters}

We now explore the control performance for hyperbolic encounters.
One challenging aspect of these simulations comes from handling the large initial
separation in the computational domain. We tested our robustness against this by
choosing different values of $D_0$ in Table \ref{tab:hyperbolic-encounters}. The
specific values of $E^*$ and $J^*$ correspond to the study of BBH scattering
angles in \spec{} \cite{Long:2025nmj} with the addition of the \texttt{q6}
run to test the control scheme beyond the equal mass, non-spinning limit.

\begin{figure}
  \centering
  \includegraphics[width=\columnwidth]{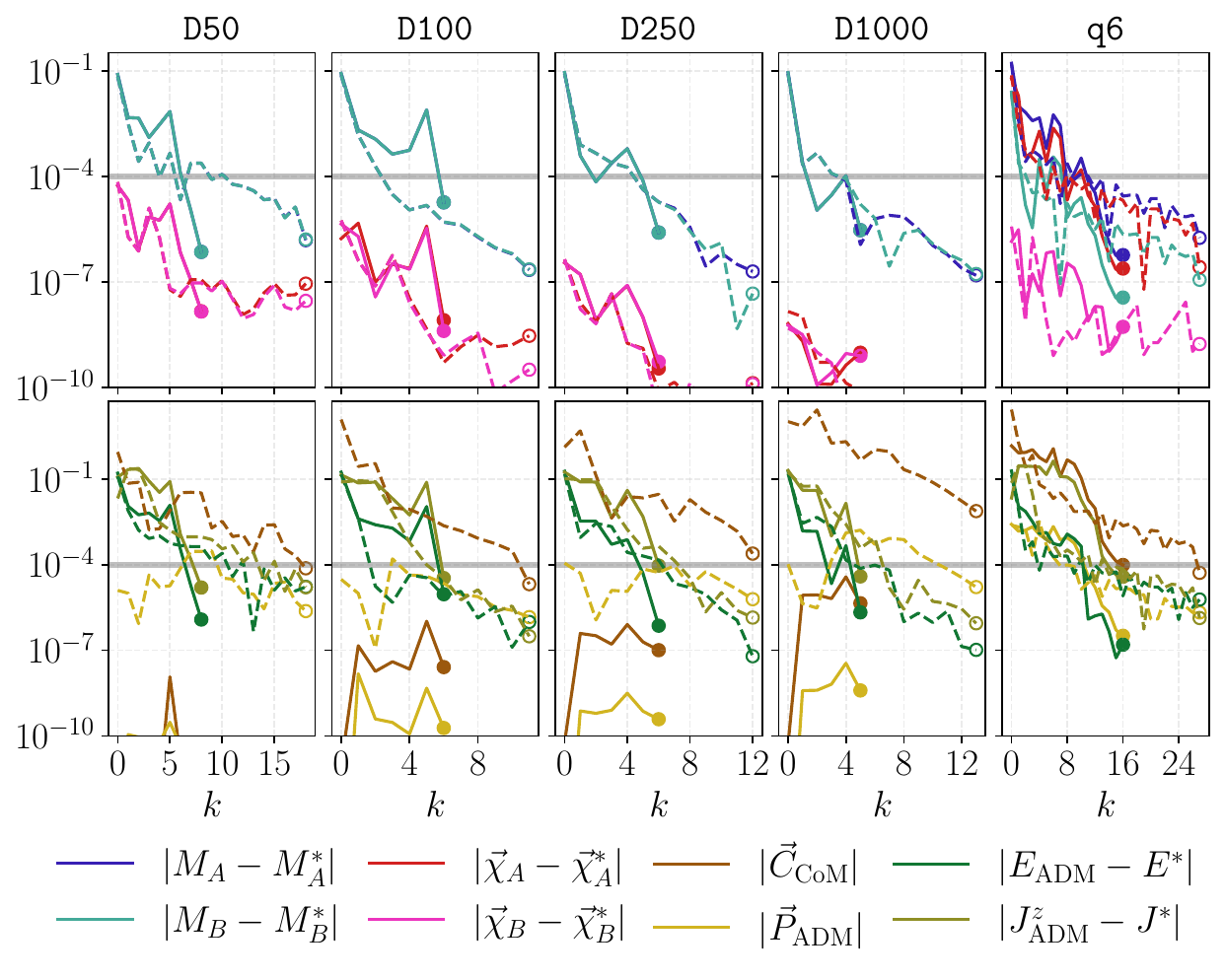}
  \caption{
    \label{fig:control-hyperbolic}
    Comparison of control iterations between \spec{} (dashed) and \spectre{} (solid) for hyperbolic encounters.
    The top panels show results for black hole masses and spins, whereas the bottom panels show the asymptotic quantities.
  }
\end{figure}

Figure \ref{fig:control-hyperbolic} compares the behavior of the control scheme
in both codes for all hyperbolic configurations in
Table \ref{tab:hyperbolic-encounters}. In all cases, \spectre{} converges in
significantly fewer iterations than \spec{} to bring
the physical parameters to their target values. In {\tt D100}, {\tt D250}, and
{\tt D1000}, we note similar patterns as before: all parameters get controlled around
the same number of iterations in the two codes, after which \spec{} waits to
have enough resolution to resolve the trivially zero center of mass.
Note that {\tt D250} and {\tt D1000} end their \spec{} control before $|\vec C_\text{CoM}| < 10^{-4}$
due to the less accurate determination of center of mass in \spec{}.
Despite also being an equal-mass case, {\tt D50} actually shows that \spectre{}
controls $J^\text{ADM}_z$ more effectively than \spec{}, indicating
improvements in the control scheme or in the computation of asymptotic quantities.

\begin{figure}
  \centering
  \includegraphics[width=\columnwidth]{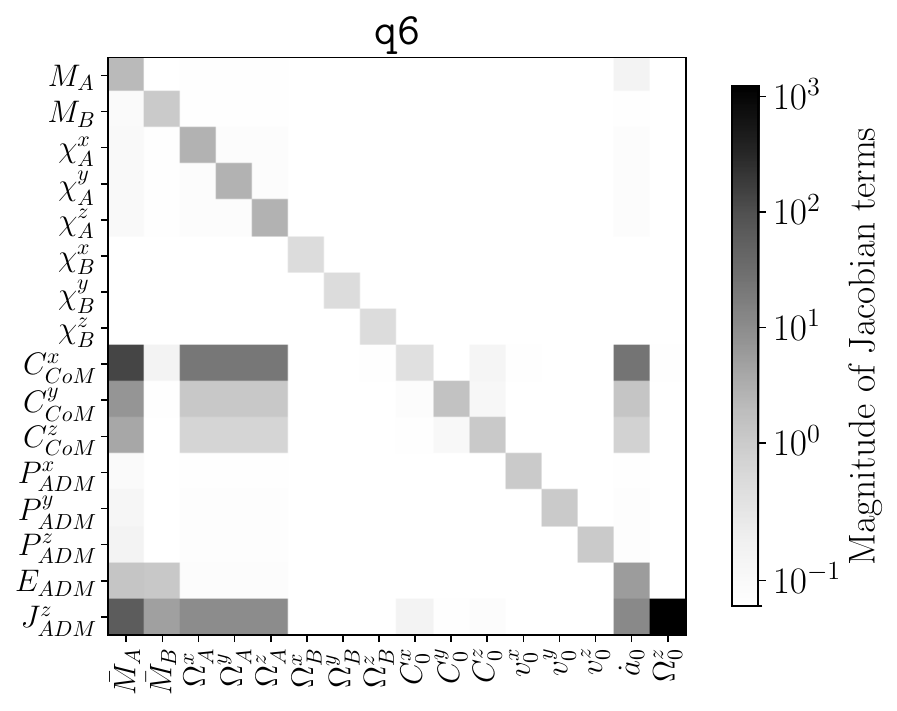}
  \caption{
    \label{fig:jacobian_hyperbolic}
    Final control Jacobian of the {\tt q6} case.
  }
\end{figure}

We choose to analyze the control Jacobian of the {\tt q6} case, which combines a
non-trivial mass ratio with the complexities of hyperbolic encounters. The final
Jacobian terms are shown in Fig.~\ref{fig:jacobian_hyperbolic}. From this, we
note that the coupling of the conformal masses with the total energy and angular
momentum from Eqs.~\eqref{eq:J-Eadm-mass-a}--\eqref{eq:J-Jadm-mass-b} are
indeed important for hyperbolic encounters. Additionally, other significant
off-diagonal terms emerge for $\vec C_\text{CoM}$ and $J^\text{ADM}_z$,
which we could try to introduce in our Jacobian initial guess if needed in the
future. These couplings could explain why using Broyden's method for all parameters
results in a more efficient control scheme.

\begin{figure}
  \centering
  \includegraphics[width=\columnwidth]{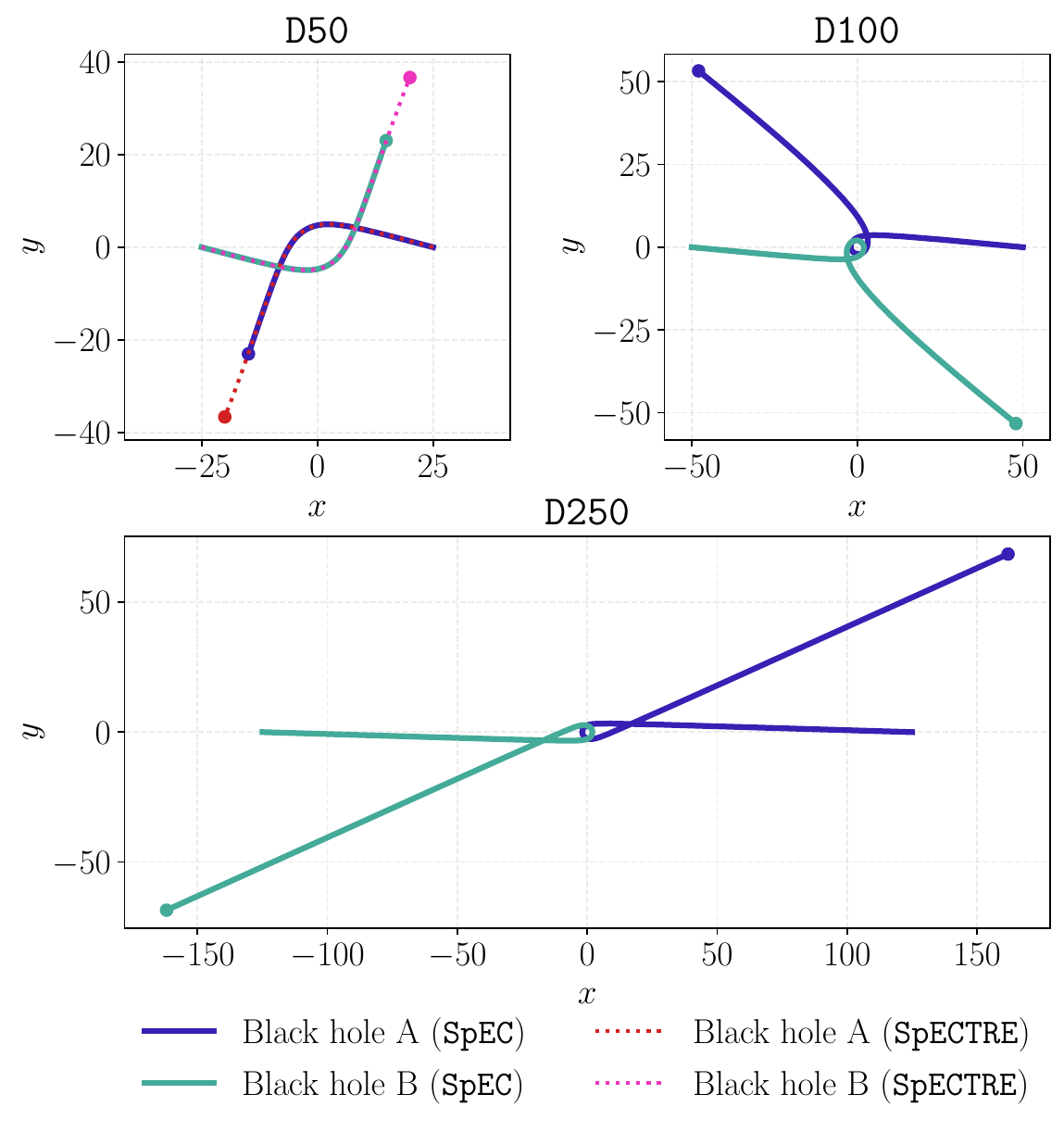}
  \caption{
    \label{fig:trajectories}
    Trajectories of hyperbolic encounters with the circles representing the
    termination points on the outgoing legs.
    {\tt D50} was evolved in both in \spec{} and \spectre{}, while {\tt D100}
    and {\tt D250} were evolved only in \spec{}.
  }
\end{figure}

The initial data of the {\tt D50}, {\tt D100}, and {\tt D250} configurations
were evolved in \spec{} in order to demonstrate the scattering trajectories enabled
by our control scheme. We also simulated the first
hyperbolic encounter in \spectre{} by evolving the smallest initial separation
case ({\tt D50}). Evolving the other hyperbolic configurations require further domain
optimizations to account for the larger separations, which we leave to future
work.
The resulting trajectories are shown in Fig.~\ref{fig:trajectories}.

\section{Conclusion}\label{sec:conclusion}

In this work, we present a parameter control scheme for
BBH initial data in the \spectre{} code.
This is the first open-source NR code with such a control procedure that
drives both horizon quantities and asymptotic properties to their target values.
By allowing the direct control of total energy and angular momentum, we enable
the construction of initial data for hyperbolic encounters.
Our scheme leverages Broyden's method in all controlled parameters to
iteratively refine the control Jacobian, allowing us to capture important
couplings between controlled parameters and free data.

We demonstrate the effectiveness of our procedure by testing it on a range of
challenging BBH configurations, including high mass ratios, extreme spins, and
large separations.
For bound orbits, our scheme achieves comparable performance
to a simpler implementation in \spec{}.
For hyperbolic encounters, we observed significant
improvement in the number of control iterations
required, highlighting the advantages of applying Broyden's method across all
parameters.

Future areas of improvement include optimizations to the
computational domain and runtime performance. For instance, it would
be ideal to implement an adaptive mesh refinement routine in the \spectre{}
control loop, similar to that currently used in \spec{} \cite{Ossokine:2015yla}.

Given the open-source nature of \spectre{}, this work makes advanced parameter
control techniques accessible to the entire numerical-relativity community.
Our implementation works directly only with XCTS excision data, but we
offer a simple interface to extrapolate the ID solution inside the excision so
that it can be used in puncture evolution codes.
With this, we aim to pave the way for more robust and accurate BBH
simulations in the era of next-generation gravitational-wave observatories.

\begin{acknowledgments}
  I.B.M. thanks Saul Teukolsky and Mark Scheel for helpful discussions, as well
  as the rest of the \spectre{} development team for the collaborative work.
  Computations were performed with the \spec{}{} and \spectre{}{}
  codes~\cite{spec, spectre} on the CaltechHPC cluster at Caltech and Urania at
  the Max Planck Institute Computing and Data Facility. The figures in this
  article were produced with \texttt{matplotlib}~\cite{matplotlib1,matplotlib2}
  and \texttt{TikZ}~\cite{tikz}.
  This work was supported in part by the Sherman Fairchild Foundation and the
  National Science Foundation under Grants No.\ PHY-2309211, No.\ PHY-2309231,
  and No.\ OAC-2209656 at Caltech.
\end{acknowledgments}

\appendix

\section{ADM integrals}\label{sec:appendix-1}

The Arnowitt-Deser-Misner (ADM) formalism defines total energy $E_\text{ADM}$,
linear momentum $\vec P_\text{ADM}$, and angular momentum $\vec J^\text{ADM}$ as
surface integrals evaluated at spatial infinity. In practice, \spectre{}
computes these asymptotic surface integrals on an outer boundary of large but
finite radius, where the computational domain is coarsest
and more prone to numerical errors. To improve accuracy of these asymptotic
quantities, we apply Gauss' theorem to transform them into volume integrals.
Since we have an excised domain in \spectre{}, this results in surface integrals over the inner boundary $S_0 = S_A \cup S_B$
and volume integrals over the entire computational domain volume $V_\infty$.

We start by writing the surface integral expression of $E_\text{ADM}$ in
Eq.~\eqref{eq:ADM-energy} in terms of conformal quantities.
From~\cite[Eq.~3.137]{BaumgarteShapiro}, we know that
\begin{equation}
  \gamma^{jk} \Gamma^i_{jk} - \gamma^{ij} \Gamma^k_{jk} =
    \psi^{-4} \Big(\bar\gamma^{jk} \bar\Gamma^i_{jk} - \bar\gamma^{ij} \bar\Gamma^k_{jk} \Big)
    - 8 \psi^{-5} \bar\nabla^i \psi.
\end{equation}
Since $dS_i = \psi^2 d\bar S_i$, Eq.~\eqref{eq:ADM-energy} becomes
\begin{equation} \label{eq:Eadm-intermediate}
  E_\text{ADM} = \frac{1}{16\pi} \oint_{S_\infty}\!\!\!  \psi^{-2} \Big(
                  \bar\gamma^{jk} \bar\Gamma^i_{jk}
                  - \bar\gamma^{ij} \bar\Gamma^k_{jk}
                  - 8 \psi^{-1} \bar\nabla^i \psi
                  \Big) \, d\bar{S}_i.
\end{equation}
Assuming that the asymptotic behavior of the conformal factor satisfies
$\psi = 1 + \mathcal{O}(1/r)$,
we can simplify Eq.~\eqref{eq:Eadm-intermediate} to \cite[Eq.~3.139]{BaumgarteShapiro}
\begin{align} \label{eq:Eadm-surface}
  E_\text{ADM} &= \frac{1}{16\pi} \oint_{S_\infty}  \Big(
                  \bar\gamma^{jk} \bar\Gamma^i_{jk}
                  - \bar\gamma^{ij} \bar\Gamma^k_{jk}
                  - 8 \bar\nabla^i \psi
                  \Big) \, d\bar{S}_i.
\end{align}
Given that the conformal metric is typically known
analytically, the first two terms in Eq.~(\ref{eq:Eadm-surface})
could in principle be directly evaluated.  However, we find
no disadvantage in treating these terms numerically, along with the last term.
Now, applying Gauss' theorem to Eq.~\eqref{eq:Eadm-surface}, we have
\begin{widetext}
\begin{align}
  E_\text{ADM}
  &= \frac{1}{16\pi} \oint_{S_0}  \Big(
    \bar\gamma^{jk} \bar\Gamma^i_{jk}
    - \bar\gamma^{ij} \bar\Gamma^k_{jk}
    - 8 \bar\nabla^i \psi
    \Big) \, d\bar{S}_i
    + \frac{1}{16\pi}
    \int_{V_\infty} \bar\nabla_i  \Big(
      \bar\gamma^{jk} \bar\Gamma^i_{jk}
      - \bar\gamma^{ij} \bar\Gamma^k_{jk}
      - 8 \bar\nabla^i \psi
    \Big) \, d\bar{V} \label{eq:Eadm-Gauss} \\
  &= \frac{1}{16\pi} \oint_{S_0}  \Big( \label{eq:Eadm-volume}
    \bar\gamma^{jk} \bar\Gamma^i_{jk}
    - \bar\gamma^{ij} \bar\Gamma^k_{jk}
    - 8 \bar\nabla^i \psi
    \Big) \, d\bar{S}_i
    \\ &\qquad \notag
    + \frac{1}{16\pi}
    \int_{V_\infty} \Big(
      \partial_i \bar\gamma^{jk} \bar\Gamma^i_{jk}
      + \bar\gamma^{jk} \partial_i \bar\Gamma^i_{jk}
      + \bar\Gamma_l \bar\gamma^{jk} \bar\Gamma^l_{jk}
      - \partial_i \bar\gamma^{ij} \bar\Gamma^k_{jk}
      - \bar\gamma^{ij} \partial_i \bar\Gamma^k_{jk}
      - \bar\Gamma_l \bar\gamma^{lj} \bar\Gamma^k_{jk}
      - 8 \bar\nabla^2 \psi
    \Big) d\bar{V},
\end{align}
\end{widetext}
where we then replace $\bar\nabla^2\psi$ by the Hamiltonian constraint
Eq.~(\ref{eq:Ham}).

The surface integral expression of $\vec P_\text{ADM}$ is given by
Eq.~\eqref{eq:ADM-linear-momentum}.
In its definition, flatness is assumed at $S_\infty$ \cite{Ossokine:2015yla}.
To avoid confusion, we introduce different notation for physical elements
$dS_i$, conformal elements $d\bar{S}_i$, and Euclidean elements $d\tilde S_j$.
Following the derivation in Ref.~\cite{Ossokine:2015yla}, we introduce a factor of
$\psi^{10}$ in the integrand, which we are allowed to do assuming that
$\psi \to 1$ at $S_\infty$.
With this, Eq.~\eqref{eq:ADM-linear-momentum} becomes
\begin{equation} \label{eq:Padm-surface}
  P_\text{ADM}^i = \frac{1}{8\pi} \oint_{S_\infty} \psi^{10} \Big(
    K^{ij} - K \gamma^{ij}
  \Big) d\tilde S_j.
\end{equation}
Applying Gauss' theorem,
\begin{equation}
  P_\text{ADM}^i
  = \frac{1}{8\pi} \oint_{S_0} P^{ij} \, d\tilde S_j + \frac{1}{8\pi} \int_{V_\infty} \partial_j P^{ij} \, d\tilde V, \label{eq:Padm-Gauss}
\end{equation}
  and using the momentum constraint
\cite[Eq.~2.128]{BaumgarteShapiro}, we have
\begin{equation}
  P_\text{ADM}^i
  = \frac{1}{8\pi} \oint_{S_0} P^{ij} \, d\tilde S_j - \frac{1}{8\pi} \int_{V_\infty} G^i \, d\tilde V, \label{eq:Padm-volume}
\end{equation}
where
\begin{align}
  P^{ij} &= \psi^{10} (K^{ij} - K \gamma^{ij}), \\
  G^i &= \bar\Gamma^i_{jk} P^{jk} + \bar\Gamma^j_{jk} P^{ik} - 2 \bar\gamma_{jk} P^{jk} \bar\gamma^{il} \partial_l(\ln \psi).
\end{align}
A similar argument is used to re-write the surface integral of $J^\text{ADM}_z$,
\begin{equation} \label{eq:Jadm-surface}
  J^\text{ADM}_z = \frac{1}{8\pi} \oint_{S_\infty} (x P^{yj} - y P^{xj}) \, d\tilde S_j,
\end{equation}
as a volume integral
\cite{Ossokine:2015yla},
\begin{align}
  J^\text{ADM}_z \label{eq:Jadm-volume}
  &= \frac{1}{8\pi} \oint_{S_0} (x P^{yj} - y P^{xj}) \, d\tilde S_j
  \\ &\qquad \notag
  - \frac{1}{8\pi} \int_{V_\infty} (x G^y - y G^x) \, d\tilde V.
\end{align}

Note that we have used Gauss' theorem in two different ways. In
Eq.~\eqref{eq:Eadm-Gauss}, we have used the conformal covariant derivative
$\bar\nabla_i$ and the conformal volume element $d\bar{V}$. In
Eq.~\eqref{eq:Padm-Gauss}, we have used the partial derivative $\partial_i$ and
the Euclidean volume element $d\tilde V$. Both approaches are equivalent
\cite{Teukolsky2016-ja}, as long as we are consistent.

\begin{figure}
  \centering
  \includegraphics[width=\columnwidth]{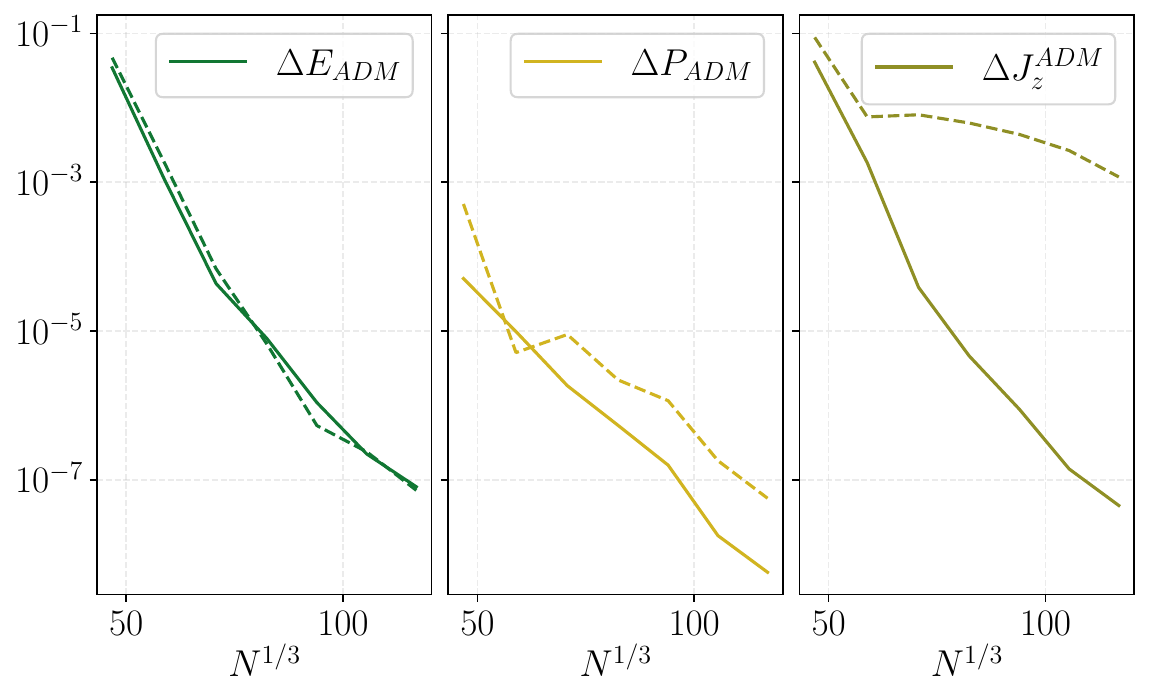}
  \caption{
    \label{fig:volume-integrals}
    Comparison of how $E_\text{ADM}$, $P_\text{ADM}$, and $J_z^\text{ADM}$ converge if computed as asymptotic surface (dashed) or volume (solid) integrals.
  }
\end{figure}

Figure \ref{fig:volume-integrals} compares the convergence of the ADM integrals
when computed using the original asymptotic surface integrals in the form of
Eqs.~\eqref{eq:Eadm-surface}, \eqref{eq:Padm-surface}, and \eqref{eq:Jadm-surface} and after application of Gauss' theorem, in the form of
Eqs.~\eqref{eq:Eadm-volume}, \eqref{eq:Padm-volume}, and \eqref{eq:Jadm-volume}.
The results shown are from the {\tt q50} test case in
Table \ref{tab:bound-orbits}.
While $E_\text{ADM}$ does not benefit from being computed as a volume integral,
we see that both $\vec P_\text{ADM}$ and $J^\text{ADM}_z$ are significantly
more precise as volume integrals. In particular, $J^\text{ADM}_z$ does not even
converge exponentially when computed as a surface integral.
Consequently, in this work, we choose to compute the momenta as in
Eq.~\eqref{eq:Padm-volume} and Eq.~\eqref{eq:Jadm-volume},
while using Eq.~\eqref{eq:Eadm-surface} to compute energy.

\spec{} follows a similar approach to compute the ADM integrals.
For $\vec P_\text{ADM}$ and $J^\text{ADM}_z$, it uses
Eq.~\eqref{eq:Padm-volume} and Eq.~\eqref{eq:Jadm-volume}, but $S_0$ is chosen
to be the inner boundary of the outer spherical shell in the computational
domain \cite{Ossokine:2015yla}.
For $E_\text{ADM}$, \spec{} expands the integrand in a $1/r$ power series,
picking up the relevant coefficients. This is analogous to the computation of
$\vec C_\text{CoM}$ in Ref.~\cite{Ossokine:2015yla}.

\section{Center-of-mass integral}\label{sec:appendix-2}

Ref.~\cite{Baskaran:2003} defines an asymptotic center of mass based on the
isometric embeddings of a two-surface $\partial\Sigma$ on a three-dimensional spatial
slice $\Sigma$ and on an auxiliary Euclidean three-space.
Here, we use $S_\infty$ as our two-surface of interest.
Let $k$ be the mean curvature of $S_\infty$ embedded in $\Sigma$.
Then, the center of mass is defined as \cite[Sec.~5]{Baskaran:2003}
\begin{equation}\label{eq:CoM-definition}
  C_\text{CoM}^i = \frac{1}{8\pi E_\text{ADM}} \oint_{S_\infty} {}^3k \, \tilde n^i \, d\Omega,
\end{equation}
where ${}^3k$ is the coefficient of the $1/r^3$ term in the asymptotic expansion
of $k$, $\tilde n^i = x^i/r$ is the Euclidean unit normal, and $d\Omega$ is the
area element of a unit sphere.  The division by $E_\text{ADM}$ in Eq.~(\ref{eq:CoM-definition}) converts from mass dipole to center of mass.
Using the sign convention from Ref.~\cite{Baskaran:2003}, the mean curvature of
$S_\infty$ is given by
\begin{align}\label{eq:k-definition}
  k = - \nabla_i n^i = - \partial_i n^i - \Gamma^i_{ij} n^j
\end{align}
where $n^i$ is the physical unit normal (i.e., normalized with respect to
$\gamma_{ij}$).
Assuming conformal flatness ($\gamma_{ij} = \psi^4 \delta_{ij}$), the
Christoffel symbols simplify to \cite[Eq.~2.7]{BaumgarteShapiro}
\begin{align}
  \Gamma^i_{ij}
  &= 2 ( \delta^i_i \partial_j \ln\psi + \delta^i_j \partial_i \ln\psi - \delta_{ij} \delta^{ik} \partial_k \ln\psi) \\
  &= \frac{6}{\psi} \partial_j \psi. \label{eq:christoffel}
\end{align}
Using Eq.~\eqref{eq:christoffel} and $n^i = \psi^{-2} \tilde n^i$,
Eq.~\eqref{eq:k-definition} becomes
\begin{equation}\label{eq:k-conformal}
  k = - \frac{2}{r} \psi^{-2} - 4 \psi^{-3} \partial_r \psi.
\end{equation}

Let us assume that
\begin{equation}\label{eq:psi-expansion}
  \psi = 1 + \frac{a}{r} + \frac{b(\theta,\phi)}{r^2} + \mathcal{O}(1/r^3),
\end{equation}
where $a$ is a constant monopole term and $b$ is an angle-dependent
dipole term. Expanding Eq.~\eqref{eq:k-conformal} in powers of $1/r$, we have
\begin{equation}
  k = - \frac{2}{r} + \frac{8a}{r^2} + \frac{12 b - 18 a^2}{r^3} + \mathcal{O}(1/r^4).
\end{equation}
From this, we identify
\begin{equation}\label{eq:k3}
  {}^3k = 12 b - 18 a^2.
\end{equation}

Using Eq.~\eqref{eq:k3} in Eq.~\eqref{eq:CoM-definition}, we find
\begin{equation}\label{eq:CoM-b}
  C_\text{CoM}^i = \frac{3}{2\pi E_\text{ADM}} \oint_{S_\infty} b(\theta,\phi) \, \tilde n^i \, d\Omega.
\end{equation}
Note that the monopole term $a$ is not angle dependent and therefore integrates
to zero.
If $\psi$ is expressed in terms of spherical harmonics, it is straightforward to
read the dipole coefficient $b$.
This is the approach taken in \spec{} \cite[Eq.~25]{Ossokine:2015yla}.
In \spectre{}, we compute $\vec C_\text{CoM}$ by directly integrating $\psi$.
From Eq.~\eqref{eq:psi-expansion}, we know that
\begin{equation}
  b = (\psi - 1) r^2 - a r + \mathcal{O}(1/r).
\end{equation}
Like before, we can ignore the monopole term $a$.
With this, Eq.~\eqref{eq:CoM-b} becomes
\begin{equation}\label{eq:CoM-final}
  C_\text{CoM}^i = \frac{3}{2\pi E_\text{ADM}} \oint_{S_\infty} (\psi - 1) \, \tilde n^i \, d\tilde A,
\end{equation}
where $d\tilde A = r^2 d\Omega$ is the Euclidean area element of $S_\infty$.
Note that we could also remove the constant term in the integrand, but we choose
to keep it to reduce round-off errors.

As a consistency test, consider a Schwarzschild black hole of mass $M$ in isotropic coordinates, which is offset from the origin by a displacement $\vec C_0$.  Such a
solution is conformally flat with conformal factor given by
\begin{align}
  \psi
  &= 1 + \frac{M}{2 |\vec x - \vec C_0|} \\
  &= 1 + \frac{M}{2 r} + \frac{M \vec{\tilde n} \cdot \vec C_0}{2 r^2} + \mathcal{O}(1/r^3).
\end{align}
Using $b = \frac{M}{2} \vec{\tilde n} \cdot \vec C_0$ in Eq.~\eqref{eq:CoM-b}, we
find indeed $\vec C_{\rm CoM}=\vec C_0$.

\begin{figure}
  \centering
  \includegraphics[width=\columnwidth]{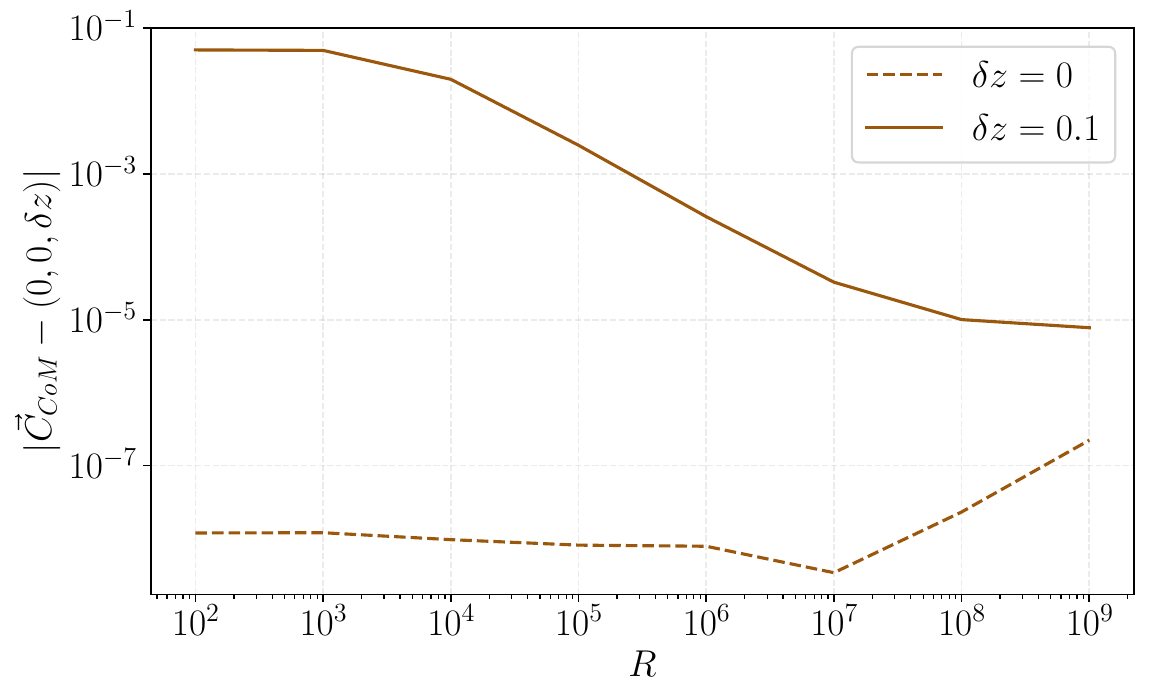}
  \caption{
    \label{fig:CoM-accuracy}
    Behavior of $\vec C_\text{CoM}$ when computed at different outer-boundary radius $R$.
  }
\end{figure}

Figure \ref{fig:CoM-accuracy} shows how Eq.~\eqref{eq:CoM-final} converges with
distance for a BBH system.
We choose to use the {\tt q1} case from Table \ref{tab:bound-orbits} because
its center of mass is trivially zero.
We also introduce a shift to the binary system $\vec C_0 = (0,0,\delta z)$, to
check convergence to a non-zero center of mass.
The unshifted case ($\delta z = 0$) shows that our error starts around
$\sim 10^{-8}$, which is the tolerance of our elliptic solver, and then grows
at large radii due to round-off errors.
The shifted case ($\delta z = 0.1$) demonstrates that we converge to the
expected center of mass as we increase the outer radius due to getting closer
to the assumed conformal flatness.
Such convergence asymptotes near $\sim 10^{-5}$, which is also likely due to round-off
errors.
Note that the center-of-mass error at $R \sim 10^5$ (outer-boundary radius used in
\spectre{}) is around $\sim 10^{-2}$ for the shifted case, but this is not a
problem because we aim to control $\vec C_\text{CoM}$ to zero.
That is, our computation of Eq.~\eqref{eq:CoM-final} becomes more accurate as we
approach $\vec C_\text{CoM} = 0$.

It is worth noting that different powers of $\psi$ could be used in the
integrand of Eq.~\eqref{eq:CoM-final}. For example, Ref.~\cite{Ossokine:2015yla}
defines the center of mass as
\begin{equation}\label{eq:CoM-Ossokine}
  C_\text{CoM}^i = \frac{3}{8 \pi E_\text{ADM}}
                     \oint_{S_\infty} \psi^4 \tilde n^i \, d\tilde A.
\end{equation}
Expanding $\psi^4$ in powers of $1/r$, we have
\begin{equation}\label{eq:psi4-expansion}
  \psi^4 = 1 + \frac{4a}{r} + \frac{4b + 6a^2}{r^2} + \mathcal{O}(1/r^3).
\end{equation}
Equation~\eqref{eq:CoM-Ossokine} follows from Eq.~\eqref{eq:CoM-b} if we ignore the
angle-independent terms in Eq.~\eqref{eq:psi4-expansion} and use
$b \approx \psi^4 r^2 / 4$.

\bibliography{References}

\end{document}